\documentclass[10pt,conference]{IEEEtran}

\usepackage{cite}
\usepackage{amsmath,amssymb,amsfonts}
\usepackage{graphicx}
\usepackage{textcomp}
\def\BibTeX{{\rm B\kern-.05em{\sc i\kern-.025em b}\kern-.08em
T\kern-.1667em\lower.7ex\hbox{E}\kern-.125emX}}
\usepackage[T1]{fontenc}
\usepackage[linesnumbered, noend, lined, ruled]{algorithm2e}
\usepackage[hidelinks]{hyperref}
\usepackage{amsthm}
\usepackage{booktabs}
\usepackage{multirow, makecell}
\usepackage[margin=10pt,font=small,labelfont=bf, labelsep=endash]{caption}
\usepackage[font=small]{subcaption}
\usepackage{listings}
\usepackage{tikz}
\usepackage{xcolor}
\usetikzlibrary{shadows, arrows, positioning, shapes, calc, automata, decorations.pathreplacing}
\usepackage{pifont}

\definecolor{color1}   {rgb}{1.0, 0.49, 0.0}
\definecolor{codeModified}{rgb}{0.82, 0.1, 0.26}
\definecolor{color2}      {rgb}{0.53, 0.66, 0.42}
\definecolor{codeAdded}{rgb}{0.0, 0.5, 1.0}
\definecolor{codeBackground1}{rgb}{0.96, 0.96, 0.96}
\definecolor{codeBackground2}{rgb}{0.52, 0.52, 0.51}
\definecolor{codeNumbers}    {rgb}{0.0, 0.0, 0.0}
\definecolor{nodeNumber}     {rgb}{0.57, 0.64, 0.69}
\definecolor{codecontainer}  {rgb}{0.98, 0.92, 0.84}
\definecolor{addEdgeColor}   {rgb}{0.0, 0.42, 0.24}
\definecolor{delEdgeColor}   {rgb}{0.9, 0.17, 0.31}
\definecolor{infoBoxColor}   {rgb}{0.74, 0.83, 0.9}

\pgfdeclarelayer{background}
\pgfdeclarelayer{foreground}
\pgfsetlayers{background,main,foreground}

\tikzset{%
  cfgNode/.style={%
    draw,
    rectangle,
    rounded corners,
    minimum height=10pt,
    minimum width=50pt,
    font=\small,
    fill=codeBackground1
  },
  cfgNodeVer1/.style={%
    cfgNode,
    fill=codeBackground2
  },
  cfgNodeVer2/.style={%
    cfgNode,
    fill=color2
  },
  cfgNodeVer3/.style={%
    cfgNode,
    fill=color1
  },
  noNode/.style={%
    draw,
    font=\footnotesize,
    circle,
    fill=nodeNumber!40,
    minimum size=5pt,
    inner sep=0.4pt
  },
  dummyNode/.style={%
    rectangle,
    rounded corners
  },
  edgeNode/.style={%
    inner sep=4pt,
    align=right,
    font=\footnotesize
  },
  addEdge/.style={%
    draw,
    thick,
    color=addEdgeColor!95,
    -latex',
    dashed
  },
  delEdge/.style={%
    draw,
    thick,
    color=delEdgeColor!95,
    -latex',
    dashed
  },
  infoBox/.style={%
    draw,
    thick,
    dashed,
    rectangle split,
    rectangle split parts=2,
    minimum width=150pt,
    minimum height=20pt,
    text centered,
    text width=150pt,
    rounded corners,
    rectangle split part fill={codeBackground1, infoBoxColor!60},
    font=\small,
  },
  multiDocument/.style={%
    shape=tape,
    rounded corners=2pt,
    draw,
    fill=codeBackground2!50,
    tape bend top=none,
    double copy shadow,
    font=\small,
  },
  multiNode/.style={%
    draw,
    rectangle,
    rounded corners,
    fill=codeBackground2!50,
    double copy shadow,
    font=\small,
  },
}

\newcommand{\nodeNo}[2]{%
  \node[noNode] at (#1.north east) {\textcolor{black}{#2}};
}

\hypersetup{pdftitle={Invariant Diffs},
  pdfsubject={Introduction to Invariant Diffs},
  pdfauthor={Ashwin Kallingal Joshy, Wei Le},
  pdfkeywords={Invariant Diff, Static Analysis, Demand-Driven}
}

\theoremstyle{definition}
\newtheorem{definition}{Definition}

\begin{document}

\title{Invariant Diffs}

\author{\IEEEauthorblockN{Ashwin Kallingal Joshy}
  \IEEEauthorblockA{\textit{Iowa State University}\\
    Ames, Iowa\\
  ashwinkj@iastate.edu}
  \and
  \IEEEauthorblockN{Wei Le}
  \IEEEauthorblockA{\textit{Iowa State University}\\
    Ames, Iowa\\
  weile@iastate.edu}
}

\maketitle

\begin{abstract}
Software development is inherently incremental. Nowadays, many software companies adopt an agile process and a shorter release cycle, where software needs to be delivered faster with quality assurances. On the other hand, the majority of existing program analysis tools still target single versions of programs and are slow and inflexible to handle changes. In the popular version control systems such as {\it git}, the program changes are still presented using source code diffs. It is hard to understand what program conditions are changed and which source code lines cause them. In this paper, we propose to compute ``invariant diffs'' to specify changes. Similar to source diffs that report common code and code churns, we define {\it version invariants\/} to represent program conditions that are common across versions, and {\it invariant churns\/} to show the changes of program conditions between versions. We designed a static {\it demand-driven}, {\it path-sensitive\/} analysis to compute and compare invariants for multiple versions of programs using {\it multiversion control flow graphs}. We report invariant diffs at the matched program points where comparing invariants are meaningful. Importantly, our analysis correlates source diffs with invariant diffs to explain what source code changes lead to the property changes. We implemented our algorithms in a tool called $H_2$ and performed experiments on 104 versions of programs. Our results show that we are able to compute invariant diffs correctly within reasonable amount of time. The version invariants can capture the common properties of program versions even constructed by different persons, and the invariant churns can specify the semantics of changes such as how a patch changed a buggy condition to a correct condition.
\end{abstract}


\section{Introduction}
As software becomes an essential part of our daily life, it is very important to be able to deliver new features, critical patches, refactoring or performance optimizations in a trustable and timely fashion. However,  it is challenging to correctly introduce a change on top of existing programs. In the past studies, researchers found that 15\%-24\% of the bug fixes are incorrect~\cite{2011:FSE:Yin}, and when programming a change, the most important information a developer wants to know is whether this change breaks any code elsewhere~\cite{2012:FSE:Tao}. Due to the lack of tool support, the developers still use source code diffs computed using levenshtein distances to understand and communicate changes in the popular version control systems such as {\it git}.

To help better understand changes and more quickly deliver reliable versions in the continuous development environment, we propose to compute ``invariant diffs'' between versions of programs to specify changes.  Like performing source code diffs which generate sections of common code and code churns, when computing invariant diffs, we generate {\it version invariants}, the program conditions that are common across versions, and {\it invariant churns}, the changes of program conditions between two versions. Program invariants have been shown important in program understanding~\cite{2001:TSE:Ernst,2000:ICSE:Ernst,2008:ICSE:Csallner}, assertion generation~\cite{2001:TCS:Nimmer,2002:ISSTA:Nimmer}, and fault localization~\cite{2008:SAC:Abreu,2013:SIGPLAN:Sahoo,2012:WODA:Alipour} for single versions of programs. We thus believe that invariant diffs can help with the similar tasks for handling changes. In fact, Menarini et al.~\cite{2017:ASE:Menarini} and Harman et al.~\cite{2018:SCAM:Harman} have pointed out the need of using change of invariants to communicate changes in code review.

In the past, Qi et al.\ proposed {\it change contracts\/}~\cite{2012:FSE:Qi}, and Lahiri et al.\ introduced {\it differential assertions\/}~\cite{2013:FSE:Lahiri}. Both of the approaches use the conditions of program variables to specify the semantics of the changes. However, to the best of our knowledge, there have not been many tools to automatically compute such information. Lo et al.\ applied dynamic analysis to infer change contracts~\cite{2014:ICSME:Le}. Person et al.\ compared symbolic constraints computed from two versions of the program to represent changes~\cite{2008:SIGSOFT:Person}. These techniques first perform analysis on a single version and then compared the results. The disadvantages are that (1) the computation is redundant when processing similar versions, (2) there are no approaches to systematically match program points of two versions, and thus the changed conditions are only reported at class or function level, (3) the analysis does not correlate source code changes with the semantics of changes, and (4) the comparison is always performed between two versions at a time, and not flexible to handle the continuous development setting, where many versions of software exist.

In this paper, we explore the definition, computation and use of invariant diffs to help understand changes and introduce reliable changes. Invariant diffs consist of two components: version invariants and invariant churns.  A version invariant specifies a common property among multiple versions. An invariant churn presents the change of the property between two versions of a program, together with the source code lines that are responsible for the property change. The property can be a condition about the value, range or typestate~\cite{typestate} of a variable, or the relationship among variables alive at the program points of interest.  We used a program representation, namely {\it multiversion control flow graph (MVICFG)\/}~\cite{2014:ICSE:Le} to automatically detect the program points that are {\it matched\/} across program versions where the computation of invariant diffs is meaningful.

We designed a static demand-driven, path-sensitive analysis on the MVICFG to compute the invariant diffs. The analysis instantiates a set of {\it invariant candidates\/} at the program point of interest using predefined templates and raise queries to ask if the invariant candidates hold along all the paths reachable from the program point on the MVICFG\@. The analysis performs a backward traversal and updates the query using the information collected at the program statements. A query is resolved when we can determine either the invariant condition holds for the paths the query has traversed, or there is a conflict with the invariant condition computed from a different path. Finally, when all the queries raised from the same invariant candidate at the same matched program points are resolved, we consolidate the invariants: if an invariant is present in more than one version, we report it as a version invariant; if different invariants are detected between two versions, we report an invariant churn.

Our techniques aim to address the four disadvantages of existing tools mentioned above. First, on the MVICFG, the nodes that are shared across multiple versions are only represented once and thus our analysis can analyze the common nodes of multiple versions simultaneously without redundant computations. Second, since demand-driven analysis only visits the nodes relevant to the query, we have the flexibility and efficiency of traversing the versions and paths based on the users' demand and compute the invariant diffs at any matched program points.  Third, our analysis is able to identify the source code lines that contribute to the computation of invariant diffs, which help explain the ``causes'' of the property changes. Importantly, although our analysis also uses templates to compute invariants, we allow the candidate invariants to contain not-yet-determined parameters specified with ``?''s. During query propagation, our analysis resolves these unknown parameters and finalizes the invariants. Fourth, given $n$ versions of a programs, we can compute version invariants that hold for any subset of versions, and compute invariant churns between any two versions of the programs.

We implemented our analysis in a tool called $H_2$, using LLVM~\cite{2004:CGO:Lattner} and Z3~\cite{2008:TACAS:Moura}. We performed the experiments on 104 versions of programs from an existing benchmark~\cite{2017:FSE:Yi} that solves the math problems. Our results show that our algorithms correctly compute version invariants and invariant churns. We experimentally demonstrated the advantages of using MVICFG to compute invariant diffs over traditional methods of computing properties for different program versions and then comparing them. We found that version invariants can represent meaningful specifications across program versions constructed by different persons, and the invariant churns capture the semantics of the bug fix by showing how the buggy conditions are changed to be the correct conditions. We also computed for each source line change, the number of invariant churns generated, and we found that the number of invariant churns are much more manageable than the number of program invariants. Thus, there is a great potential to use invariant churns to assist code reviews for changes.

In summary, the contributions of this paper include:
\begin{enumerate}
    \item The concept of invariant diffs, including version invariants and invariant churns,
    \item Demand-driven, path-sensitive analysis to statically compute the invariant diffs, and
    \item $H_2$, a framework and prototype tool, and the experiments that demonstrate our algorithms are correct, and the version invariants and invariant churns have real software engineering values.
\end{enumerate}

The rest of the paper is organized as follows: Section~\ref{sec:Overview} presents the key ideas of the work, Section~\ref{sec:Alg} explains our approaches of computing invariant diffs. In Section~\ref{sec:Evaluation}, we describe our implementation of $H_2$ and present the experimental results. Section~\ref{sec:Related} discusses the related work, followed by the conclusions and future work in Section~\ref{sec:Conclusion}.

\section{Key Ideas}\label{sec:Overview}
\begin{figure*}
  \centering
  \resizebox{0.22\textwidth}{140pt}{%
    \begin{subfigure}[b]{120pt}
\begin{lstlisting}[frame=none]
foo();
a = 2;
cin >> a1;
b = a + 1;
c = 2;
d = c + a1;
... !\bigskip!



\end{lstlisting}
      \caption{Version 1}\label{fig:Ver1}
    \end{subfigure}%
  }\qquad
  \resizebox{0.24\textwidth}{140pt}{%
    \begin{subfigure}[b]{120pt}
\begin{lstlisting}[frame=none]
foo();
a = 2;
cin >> a1;
b = a + 1;
!\color{codeAdded}{if (b > a1) \{}!
  !\color{codeAdded}{a1++;}!
!\color{codeAdded}\}!
!\color{codeModified}{c = b - 1;}!
d = c + a1;
...
\end{lstlisting}
      \caption{Version 2}\label{fig:Ver2}
    \end{subfigure}%
  }\qquad
  \resizebox{0.28\textwidth}{140pt}{%
    \begin{subfigure}[b]{120pt}
\begin{lstlisting}[frame=none]
foo();
a = 2;
cin >> a1;
b = a + 1;
if (b > a1) {
  a1++;
  !\color{codeAdded}{b++;}!
!\color{codeAdded}{\} else \{}!
!\color{codeAdded}{  a1 = 2;}!
!\color{codeAdded}{  b = 4;}!
}
!\color{codeModified}{c = b - 2;}!
d = c + a1;
...
\end{lstlisting}
      \caption{Version 3}\label{fig:Ver3}
    \end{subfigure}%
  }\qquad
\caption{Program Versions and Source Code Diffs}~\label{fig:SourceCode}
\end{figure*}

In this section, we use a simple example to explain the key ideas of our work. In Figure~\ref{fig:SourceCode}, we present 3 versions of code snippets. Its MVICFG is given in Figure~\ref{fig:MVICFG}. In the graph, the statements common across the 3 versions, including nodes 0--3, 5 and 14, have been presented only once. The edges in the MVICFG are annotated with version information. For example, the edge $\left<3,6\right>$ indicates that node~6 belongs to versions 2 and 3, and the edge $\left<2,3\right>$ indicates that node~3 is shared across versions 1--3. We mark {\it T\/} and {\it F\/} to specify the true and false branches.

\begin{figure}
  \centering
  \begin{tikzpicture}[->, > = stealth', shorten > = 1pt, auto, node distance = 40pt]
  \node [cfgNode]   (start)   []                                 {foo};
  \node [cfgNode]     (a)     [below = 10pt of start]            {a = 2};
  \node [cfgNode]     (inp)   [below = 10pt of a]                {cin >> a1};
  \node [cfgNode]     (b)     [below = 10pt of inp]              {b = a + 1};
  \node [cfgNodeVer2] (if)    [below right = 10pt and 65pt of b] {if (b > a1)};
  \node [cfgNodeVer2] (a1)    [below left = 20pt and 5pt of if]  {a1++};
  \node [cfgNodeVer3] (b1)    [below right = 10pt and 5 ptof a1] {b++};
  \node [cfgNodeVer3] (b1)    [below right = 10pt and 5 ptof a1] {b++};
  \node [cfgNodeVer3] (a2)    [below right = 10pt and 5pt of if] {a1 = 2};
  \node [cfgNodeVer3] (b2)    [below = 10pt of a2]               {b = 4};
  \node [cfgNodeVer2] (c2)    [below left = 15pt and 5pt of b1]  {c = b - 1};
  \node [cfgNodeVer3] (c3)    [right = 30pt of c2]               {c = b - 2};
  \node [cfgNodeVer1] (c)     [left = 10pt of c2]                {c = 2};
  \node [cfgNode]     (d)     [below = 25pt of c]                {d = c + a1};
  \node [cfgNode]   (stop)    [below = 10pt of d]                {$\dots$};

  \nodeNo {start} {0};
  \nodeNo {a}     {1};
  \nodeNo {inp}   {2};
  \nodeNo {b}     {3};
  \nodeNo {c}     {4};
  \nodeNo {d}     {5};
  \nodeNo {if}    {6};
  \nodeNo {a1}    {7};
  \nodeNo {c2}    {8};
  \nodeNo {a2}    {10};
  \nodeNo {b1}    {11};
  \nodeNo {b2}    {12};
  \nodeNo {c3}    {13};
  \nodeNo {stop}  {14};

  \path
    (start.south) edge [->] node [edgeNode, left]                       {<1--3>}                                                  (a.north)
    (a.south)     edge [->] node [edgeNode, left]                       {<1--3>}                                                  (inp.north)
    (inp.south)   edge [->] node [edgeNode, left]                       {<1--3>}                                                  (b.north)
    (b.south)     edge [->] node [edgeNode, left]                       {<1>}                                                     (c.north)
    (c.south)     edge [->] node [edgeNode, left]                       {<1>}                                                     (d.north)
    (d.south)     edge [->] node [edgeNode, left]                       {<1--3>}                                                  (stop.north)
    (b.south)     edge [->] node [edgeNode, right, yshift=5pt, pos=0.4] {<2--3>}                                                  (if.north)
    (if.south)    edge [->] node [edgeNode, right, yshift=0pt, pos=0.6]           {<2--3>} node [edgeNode, right, yshift=-3pt, pos=0.2] {T} (a1.north)
    (a1.south)    edge [->] node [edgeNode, right]                      {<2>}                                                     (c2.north)
    (c2.south)    edge [->] node [edgeNode, left, yshift=5pt]           {<2>}                                                     (d.north)
    (if.south)    edge [->] node [edgeNode, right, yshift=5pt]          {<3>}    node [edgeNode, left, yshift=-3pt, pos=0.4]  {F} (a2.north)
    (a1.south)    edge [->] node [edgeNode, right, yshift=5pt]          {<3>}                                                     (b1.north)
    (b1.south)    edge [->] node [edgeNode, right]                      {<3>}                                                     (c3.north)
    (a2.south)    edge [->] node [edgeNode, right]                      {<3>}                                                     (b2.north)
    (b2.south)    edge [->] node [edgeNode, right, pos=0.35]            {<3>}                                                     (c3.north)
    (c3.south)    edge [->] node [edgeNode, right,yshift=-5pt]          {<3>}                                                     (d.north);

    \draw[->, rounded corners=6mm] (if.south) -- ($(if.south) + (-65pt,-5pt)$) node[edgeNode, left, yshift=5pt, pos=0.5]{F} -- ($(a1.west) + (-15pt, 10pt)$) -- node[edgeNode, left, yshift=5pt, pos=0.5]{<2>} (c2.north);
\end{tikzpicture}
  \caption{MVICFG for Versions 1--3 shown in Figure~\ref{fig:SourceCode}}\label{fig:MVICFG}
\end{figure}

\subsection{Defining Version Invariants and Invariant Churns}~\label{subsec:Def}
Program invariants are the conditions held true at a program point of interest along all the program paths of a program. Given a set of program versions, if an invariant remains the same at a program point common for all these program versions, we call such an invariant the {\it version invariant}. We call such program point the {\it matched program point}. The matched program points are the places where no source code diffs are reported. We used an existing tool MVICFG to automatically find these program points on the control flow graphs of the program versions.

\begin{definition}~\label{def:VerInv}
  A \textit{version invariant} for $n$ program versions is a condition held true for all the versions at their matched program point.
\end{definition}

At node 5 in Figure~\ref{fig:MVICFG},  there is a version invariant $c = 2$. In version 1, $c$ is assigned to 2 at node~4. In version 2, along paths $\left<0-3, 6,(7), 8\right>$,  $b$ gets 3 at node~3, and then $c$ gets 2 at node 8. Similarly, in version 3, $b$ is updated to be 4 along both the true branch $\left<6,7,11\right>$, and the false branch $\left<6,10,12\right>$; therefore, $c$ gets 2 at node~13. Since the invariant $c = 2$ is present in the 3 versions of the program at node 5, it is a version invariant at node~5.

The second key concept is \textit{invariant churn}. Similar to its namesake code churn, an invariant churn records the invariants that are added and deleted between two versions of a program at their matched program point.

\begin{definition}~\label{def:InvChurn} An \textit{invariant churn} between two program versions is a 3--tuple, \[(mp, \left\{\pm inv_i\right\},  \left\{\left\{\pm code_j\right\},S\right\})\] where,
  \begin{itemize}
    \item $mp$ is the matched program point;
    \item $\pm inv_i$ is the pair of changed invariants at the matched program point; $+inv_i$ denotes the invariants hold only in the second version but not in the first version, whereas $-inv_i$ denotes the invariants only hold for the first version but not in the second version; and
    \item $\left\{\pm code_j\right\}$ reports the source code churns that are responsible for the invariant changes; $+code_j$ denotes the lines newly added in the second version, and $-code_j$ denotes the lines only existing in the first but not in the second version. $S$ contains a set of source code lines shared by the two versions but contribute to the invariant changes.
  \end{itemize}
\end{definition}

In Figure~\ref{fig:MVICFG}, there is an invariant churn between versions~2 and 3: $\big(5, \left\{-[b = 3], +[b = 4]\right\}, \left\{\left\{+11,+12\right\},\left\{1,3\right\}\right\}\big)$. The churn is reported at the matched point node~5. The invariant $b=3$ in version 2 is changed to $b=4$ in version 3. Nodes 11 and 12 newly added in version 3, together with the shared nodes 1 and 3, are reported as affecting this invariant change. There is also an invariant churn $\big(5, \left\{- [d = 2 + input_1]\right\}, \left\{ \left\{+7,+8,-4,\right\},\left\{2,5\right\}\right\}\big)$ reported at node~5 between versions 1 and 2.  It indicates that the second version removes the invariant $d = 2 + input_1$, where $input_1$ represents the input value for $a1$ at node 2. Specifically, in version 1, there is only one path $\left<1-5\right>$, where $d$ is computed by $c+a1$ at node~5. Thus the invariant at node~5 is $d = 2 + input_1$. In the second version, $a1$ incremented once in the true branch at node~7 and left unmodified in the false branch.  Hence, $d$ can be either $2 + input_1$ or $2 + (input_1 +1)$. The previous invariant no longer hold.

\subsection{Computing Invariant Diffs via Demand-Driven Analysis}
We developed a static demand-driven, path-sensitive symbolic analysis on top of MVICFG to compute version invariants and invariant churns. To detect the version invariant $c=2$ at node~5 given in Section~\ref{subsec:Def}, we use a set of invariant templates~\cite{2001:TSE:Ernst} to construct a set of queries at node~5, one of which is $c=?$, asking if a {\it constant invariant\/} regarding $c$ potentially exists, and if so, what is the value. The query is propagated backwards along all the paths for all the versions of interest reachable from node 5 on the MVICFG\@.

At node~4, $c$ is assigned to 2, and the invariant $c=2$ is discovered for version 1. At node~8, $c$ is assigned to $b - 1$ and we update the query to $c=b-1$. This query is simultaneously  propagated to the true branch $\left<7,6\right>$, and the false branch $\left<6\right>$. We gain no additional information about $c$ along the branches, and thus the two copies of the same query are merged at node~6. The query continues propagating backwards. When it arrives at node~3, the query is updated to $c=a?$, which then is resolved at node~1, reporting $c=2$ as an invariant for version 2. Similarly, there are two copies of queries propagated to the branches at node13. Along path $\left<13,12\right>$, the query is resolved at node~12, indicating $c=2$. Along path $\left<13,11,7,6,3-1\right>$, we report $c=2$. We therefore can conclude $c=2$ is a version invariant valid for versions 1--3.

To generate the invariant churn given in Section~\ref{subsec:Def}, $\big(5, \left\{-[b = 3], +[b = 4]\right\}, \left\{\left\{+11,+12\right\},\left\{1,3\right\}\right\}\big)$. We used the query $b=?$. To compute the invariant churn between versions 2 and 3, we propagate the query only along the paths marked with versions 2 and 3. Using a similar query propagation approach, we can resolve $b=3$ for version 2, and $b=4$ for version 3 at node 1. We also record the nodes that update the query, including the shared nodes~1 and 3 and the added nodes 11 and 12. In another example where we obtain $\big(5, \left\{- [d = 2 + input_1]\right\}, \left\{ \left\{+7,+8,-4,\right\},\left\{2,5\right\}\right\}\big)$ between versions 1 and 2, the query $d=?$ gets resolved at node~2 as $d = 2 + input_1$ along $\left<5-2\right>$, but along $\left<8,7,6\right>$ and  $\left<8,6\right>$, the query becomes $d=b+a1$ and $d=b+a1-1$ respectively. We thus terminate the query propagation at node~6, reporting no invariants discovered for version 2.

The above examples show that using the demand-driven analysis, we only visited the code relevant to the queries. In all of the above cases, the query is resolved and the analysis is terminated without traversing the function {\tt foo}, which can be quite complicated. Utilizing MVICFG, we can analyze multiple versions simultaneously and visit the shared nodes only once to determine the invariants for multiple versions. For example, the query $d=?$ is updated at node~5 only once when detecting invariants for versions 1--3. Using MVICFG, we can also easily compare in-progress invariants of different versions. For example, if we aim to determine whether $a1=?$ holds for versions 1 and 2. When we find that the queries from $\left<5-3\right>$ and $\left<8-6\right>$ would surely lead to different results, we can terminate the analysis early and report there is no constant version invariant regarding $a1$ between versions 1 and 2.

In addition to the constant invariant template, our analysis also supports the {\it inequality template}. For example, at node~5 in Figure~\ref{fig:MVICFG}, we can construct a query $b>a1?$ to ask if the invariant $b>a1$ holds between versions 1 and 3. Along path $\left<5-3\right>$, the query is resolved at node~3 and reports that $b>a1$ holds for version 1. For version 3, along path $\left<5,13,11,7,6\right>$, we update the query to be $b+1>a1?$ at node~11, and at node~6, using the condition $b>a1$ from the branch node, we determine $b+1>a1$ holds for the path. Along  $\left<5,13,12,10\right>$, we obtain $4>2$, which is always true. Thus, $b>a1$ is a version invariant for versions 1 and 3. We also have the invariant templates for computing arithmetic relations between variables, such as $a=?b+?c$ (``?'' here represents not-yet-solved parameters). Our analysis is able to determine if such invariants exist and what are the values for the parameters (details in Section 3).

Our examples indicate that we used the same query to compute both version invariants and invariant churns. Our approach computes meaningful invariant churns because (1) the invariants in the churns are regarding the same variables given by the same invariant candidate at the matched program point, (2) the paths of two versions at least share the matched program points, and may share more nodes, and (3) the analysis is able to track how the differences in the paths of two versions change the presence of the invariants and/or generate different values for ``?''s which also lead to different invariants.

\subsection{The Use Scenarios}
\paragraph{Help understand code changes} Invariant churns provide how an invariant in a version is changed, what source code changes ``cause'' the property change, and which part of the non-changed code is also relevant. For an invariant change of interest, developers can inquiry which code churns are relevant, and similarly, for a particular code churn of interest, which conditions of a program are updated as a result. Our techniques can be integrated in the continuous development environment, and work with developers interactively for such inquires about changes. They can also be used to generate invariant churns together with code churns as a commit message when the developers check in their code to the version control repositories.

\paragraph{Generate assertions} A version invariant shows what has not changed, and it provides a specification on what are the common conditions across the versions. Such specification can be used by program analysis and testing tools, and it also can be asserted in the program to prevent future bugs. When introducing changes to a software repository, we typically do not know which conditions are important and thus should not be changed. As a result, we may repeatedly introduce bugs to break the ``must-obey-rules''.

Compared to invariants generated based on one version, we believe that version invariants are stronger and less dependent on specific implementations tied to a version. We also can use version invariants together with invariant churns to further help select important specifications. Suppose a change is a patch, and the removed invariant in the invariant churn is a version invariant. It may reflect that the bug has existed in the code for a long time, and thus the newly added invariant represent the correct conditions and can be used as assertions to prevent the similar mistakes in the future. On the other hand, suppose a change introduces a bug, and the removed invariant is a version invariant. It may reflect that a correct condition held for a long time is broken, and thus the version invariant can be used as assertions. In a realistic scenario, there can be additional changes checked in with a patch, but as long as developers can point out which code churn is related to the patch, we can get the invariants relevant to the fix for generating assertions.

\paragraph{Help localize the fault} To diagnose a regression bug introduced in a change, we can inspect the invariant churns. If the error conditions can be found in the invariant churns related to the buggy program version, we can pinpoint the source line changes associated with the invariant churn as the root cause. We should prioritize the invariant churns where the removed invariants are the version invariants, as the version invariants likely represent the correct behaviors from the past versions.

\section{The Algorithms of Computing Invariant Diffs}\label{sec:Alg}
In this section, we present how to compute invariant diffs. We first present a high level design of our analysis, including how to reduce the problem of computing invariants in versions to a demand-driven analysis, and what are the key components used for analysis. We then give details of the algorithm and explain why it is correct.

\subsection{Formulating the Problem to Demand-Driven Analysis}
Static demand-driven analysis has been shown feasible, efficient and flexible for detecting bugs and verifying patches~\cite{2008:FSE:Le,2014:ICSE:Le,2007:PASTE:Le}. Instead of exhaustively traversing all the paths from the entry of the program to determine a property, demand-driven analysis formulates the demand using queries, and traverses the paths relevant to the queries to determine the property. For applying demand-driven analysis to detect version invariants and invariant churns, we used the MVICFG as a program representation. The MVICFG makes available the common and the changed paths between versions as well as the matched program points across versions. Based on the demand, we then can drive the analysis only to the versions and changes of interest available on MVICFG\@.

We predefined a set of templates that specify the invariants, following the format $y\sim g(x_1, x_2, \ldots)$, where (1) $x_1$, $x_2$, $\ldots$, $y$ are the variables of the program alive at the program point where the invariants are computed, (2) $\sim \in \left\{>, <, =, \neq, \geq, \leq \right\}$, (3) $g(x_1, x_2, \ldots)$ is a polynomial expression that specifies the relations between variables, and (4) ``$?$'' can be used in $g$ to represent ``to-be-determined'' parameters for the variables. We construct a query at the matched program point of interest given by the user. Typically, for invariants, the program points in the loop, at the entry or exit of a procedure are often meaningful. The query contains an {\it invariant candidate\/} instantiated from the template such as $c=?$, $c=?a+?b$ or $b>a1$. To determine if an invariant exists, we perform a path-sensitive analysis to determine whether along all paths, the invariant candidate hold. If the invariant candidate includes ``?''s, our analysis will invoke the constraint solver over a conjunction set of invariant candidates collected from different paths. The solver answers what should be the values of ``?''s to make the invariant candidate hold for all the paths. If such a value does not exist, there are no invariants of this format at this program point.

The demand-driven analysis performs a backward traversal along the paths on the MVICFG reachable from where the query is raised, and collects the information from the program statements to determine the query resolution. The query can be resolved to {\it yes}, meaning the invariant candidate holds for the paths the query has traversed, or {\it no\/} when along different paths, we reach different conditions, and thus invariant candidate does not hold. Once raised, a query can generate many instances or copies along different paths. When all the instances are resolved, we consolidate the resolutions to generate version invariants and invariant churns regarding the query. The analysis is able to record which program statements updated the query and thus contribute to the computation of the invariants. These statements are reported with invariant churns to help explain what causes the invariant changes.

\subsection{The Components of $H_2$}
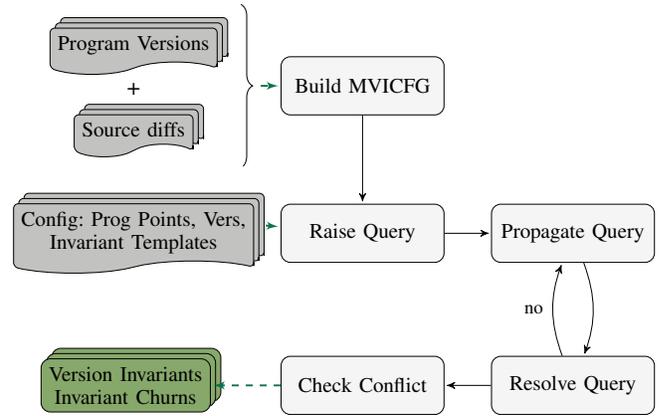
\begin{figure}
  \centering
  \resizebox{\columnwidth}{!}{%
    \begingroup
  \tikzset{%
    cfgNode/.style={%
      draw,
      rectangle,
      rounded corners,
      minimum height=25pt,
      minimum width=70pt,
      font=\small,
      fill=codeBackground1
  }}
  \begin{tikzpicture}[->, > = stealth', shorten > = 1pt, auto, node distance = 40pt]
    \node[multiDocument] (cVers)                        {Program Versions};
    \node[]              (plus)  [below = 5pt of cVers] {+};
    \node[multiDocument] (diff)  [below = 5pt of plus]  {Source diffs};
    \node[multiDocument, align=center]     (pp)    [below = 25pt of diff] {$\text{Config: Prog Points, Vers,}$ \\ Invariant Templates$$};
    \draw [-,decoration={brace, amplitude=5pt, raise=10pt}, decorate] ([shift={(0pt,10pt)}]cVers.north east) -- ([shift={(0pt,-60pt)}]cVers.north east) node[midway, xshift=5pt] (midway) {};

    \node[cfgNode]   (mvicfg)     [right = 15pt of midway]   {Build MVICFG};
    \node[cfgNode]   (raiseQ)     [below = 38pt of mvicfg]   {Raise Query};
    \node[cfgNode]   (propQ)      [right = 20pt of raiseQ]   {Propagate Query};
    \node[cfgNode]   (resolveQ)   [below = 40pt of propQ]    {Resolve Query};
    \node[cfgNode]   (conflict)   [left = 20pt of resolveQ]  {Check Conflict};

    \node[multiNode, fill=color2, align=center] (out)  [left = 30pt of conflict]  {Version Invariants \\ Invariant Churns};

    \path
      (midway)+(10pt,0pt) edge [addEdge] (mvicfg)
      (mvicfg)            edge (raiseQ)
      (pp.east)+(5pt,3pt) edge [addEdge] (raiseQ)
      (raiseQ)            edge (propQ)
      (propQ)             edge [bend left=20] (resolveQ)
      (resolveQ)          edge [bend left=20] node[edgeNode] {no} (propQ)
      (resolveQ)          edge (conflict)
      (conflict)          edge [addEdge] (out.east);
  \end{tikzpicture}
  \endgroup
  }
  \caption{The Components of $H_2$}\label{overVwHydrogen}
\end{figure}

We developed a framework called $H_2$ to compute invariant diffs. As shown in Figure~\ref{overVwHydrogen}, we take the program versions and their corresponding source code churns as input to generate the MVICFG~\cite{2014:ICSE:Le}. Dependent on the application scenarios, the users may specify the program versions, invariant templates or program points of interest for computing invariant diffs. If needed, $H_2$ can also compute invariant diffs for all the matched program points based on all the program variables alive at the program points, considering all the versions. The output of $H_2$ includes version invariants and invariant churns.

$H_2$ consists of a set of components similar to other demand-driven analyses~\cite{2005:OOPSLA:Sridharan,2001:PLDI:Heintze,2011:ISSTA:Yan,2008:POPL:Zheng,2008:FSE:Le, 2007:PASTE:Le, 2014:ICSE:Le}. The {\it Raise Query\/} component constructs a query. It generates the invariant candidate from the template, and it also initializes the other fields in the query that will be used in the query propagation. The {\it Propagate Query\/} component encapsulates the demand-driven query propagation algorithms, including how to advance the queries through the edges marked with versions and how to merge queries at branch and shared nodes. When a query arrives at a new node and gets updated, the {\it Resolve Query\/} component determines whether the invariant candidates hold for the paths traversed. If the query is not yet resolved, {\it Propagate Query\/} continuously advances the query on the MVICFG to collect more information. If the answer is yes, the {\it Check Conflict\/} component compares the invariants computed from this query to invariants from other paths and other versions for generating version invariants and invariant churns.

\subsection{The Algorithm}
In Algorithm~\ref{alg:Overview}, we present further details of computing invariant diffs. The algorithm takes 5 inputs; (1) $n$ versions of a program ($p_1,p_2,\dots,p_n$), (2) the source diffs ($d_1,d_2,\dots,d_{n-1}$), where $d_i$ is the diff between $p_i$ and $p_{i+1}$, (3) a set of program points of interest ({\bf\ttfamily P}), (4) a set of program versions ({\bf\ttfamily V}) for which the users want to compute invariant diffs, and (5) a set of invariant templates of interest ({\bf\ttfamily InvT}). The last three inputs are the configurations the user can specify to interactively use our tool. The output of the algorithm is the version invariants at {\bf\ttfamily P} for any versions in {\bf\ttfamily V} and the invariant churns at {\bf\ttfamily P} between any two versions in {\bf\ttfamily V}.

\begin{algorithm}[h!tb]
  \SetAlgoVlined{}
  \DontPrintSemicolon{}
  \SetKwInOut{Input}{Input}\SetKwInOut{Output}{Output}
  \SetKwProg{Fn}{\bf Function }{}{end}
  \SetKwData{InvC}{InvT}
  \SetKwData{P}{P}
  \SetKwData{p}{p}
  \SetKwData{V}{V}
  \SetKwData{MVICFG}{mvicfg}
  \SetKwData{MP}{MatchedPoint}
  \SetKwData{Vars}{Vars}
  \SetKwData{var}{var}
  \SetKwData{can}{t}
  \SetKwData{q}{q}
  \SetKwData{WorkList}{WorkList}
  \SetKwData{Inv}{Inv}
  \SetKwData{VI}{VI}
  \SetKwData{IC}{IC}
  \SetKwData{False}{False}
  \SetKwData{True}{True}
  \SetKwData{i}{i}
  \SetKwData{t}{t}
  \SetKwData{invM}{$i_m$}
  \SetKwData{invN}{$i_n$}
  \SetKwData{vI}{v$_i$}
  \SetKwData{vIp}{v$_{i+1}$}
  \SetKwData{lines}{lines}

  \SetKwFunction{GenMVICFG}{BuildMVICFG}
  \SetKwFunction{AliveVars}{GetVars}
  \SetKwFunction{RaiseQuery}{RaiseQuery}
  \SetKwFunction{PropagateQuery}{PropagateQuery}
  \SetKwFunction{ResolveQuery}{ResolveQuery}
  \SetKwFunction{Resolved}{Resolved}
  \SetKwFunction{Conflicting}{CheckConflict}
  \SetKwFunction{AddInv}{UpdateInV}
  \SetKwFunction{UpdInv}{RemoveInv}
  \SetKwFunction{UpdWLQueries}{RemoveWorklistQueries}
  \SetKwFunction{MergeOrTerminateQ}{MergeOrTerminateQ}
  \SetKwFunction{VisitedNodeBefore}{VisitedNodeBefore}
  \SetKwFunction{IntChurn}{IntChurn}
  \SetKwFunction{ComputingIC}{ComputingIC}
  \SetKwFunction{ComputingVI}{ComputingVI}
  \SetKwFunction{AddInvChurn}{AddInvChurn}

  \Input{Program versions ($p_1, p_2, \dots, p_n$)\\
    Source diffs ($d_1, d_2, \dots, d_{n-1}$)\\
    Program points of interest (\P)\\
    Program versions of interest (\V)\\
    Invariant templates of interest (\InvC)}
  \Output{Version Invariants, Invariant Churns}
  \BlankLine{}
  \MVICFG$\leftarrow$\GenMVICFG{$p_1,\dots, p_n$,$d_1,\dots,d_{n-1}$}\;\label{alg:OverView:GenMVICFG}
  \WorkList$\leftarrow\emptyset$; \Inv$\leftarrow\emptyset$\;\label{alg:OverView:Intialize}
  \ForEach{$\p \in \P$}{\label{alg:OverView:ForEachLoop}
    \If{\p\ is a \MP\ over \V}{\label{alg:OverView:IfCond}
      \Vars$\leftarrow$\AliveVars{\p,\V,\MVICFG}\;\label{alg:OverView:Vars}
      \ForAll{$\can \in \InvC$}{\label{alg:OverView:ForAllLoop}
        \q$\leftarrow$\RaiseQuery{\p,\V,\can,\Vars}\;\label{alg:OverView:RaiseQ}
        Add \q\ to \WorkList\;\label{alg:OverView:AddWorkList}
      }
    }
  }
  \While{\WorkList\ is not empty}{\label{alg:OverView:WhileLoop}
    Remove \q\ from \WorkList\;\label{alg:OverView:getQ}
    \If{$\neg$\ResolveQuery{\q,\Inv,\WorkList}}{\label{alg:OverView:ResolveQ}
      \PropagateQuery{\q,\MVICFG}\label{alg:OverView:PropQ}
    }
  }
  \ComputingVI{\P,\V,\Inv}\;\label{alg:Overview:CompVI}
  \ComputingIC{\P,\Inv,\IC}\;\label{alg:Overview:CompIC}

  \BlankLine%
  \Fn{\ResolveQuery{\q,\Inv,\WorkList}}{\label{alg:ResQ}
    \uIf{\Resolved{\q}}{\label{alg:ResolveQuery:IfResolved}
      \uIf{$\neg$\Conflicting{\q, \Inv}}{\label{alg:ResolveQuery:IfConflicting}
        \AddInv{\q, \Inv}\label{alg:ResolveQuery:AddInv}
      }\Else{\label{alg:ResolveQuery:ElseConflicting}
        \UpdInv{\q, \Inv}\;\label{alg:ResolveQuery:UpdInv}
        \UpdWLQueries{\q, \WorkList}\label{alg:ResolveQuery:UpdWL}
      }
      \Return{\True}\label{alg:ResolveQuery:RetTrue1}
    }\ElseIf{\VisitedNodeBefore{\q}}{\label{alg:ResolveQuery:VisitedNode}
      \MergeOrTerminateQ{\q, \WorkList}\;\label{alg:ResolveQuery:MergeT}
      \Return{\True}\label{alg:ResolveQuery:RetTrue2}
    }
    \Return{\False}\label{alg:ResolveQuery:RetFalse}
  }
  \caption{Computing Invariant Diffs}\label{alg:Overview}
\end{algorithm}

At line~\ref{alg:OverView:GenMVICFG}, we first generate the MVICFG according to~\cite{2014:ICSE:Le}. The source code churns $d_i$ can be generated using tools like UNIX diff~\footnote{\url{https://www.gnu.org/software/diffutils}}. At line~2, we initialize {\bf \ttfamily WorkList}, used for propagating the queries, and {\bf \ttfamily Inv}, used for storing the {\it invariants-in-progress\/} generated by the queries --- these are the invariants collected when the queries are resolved.

At line~\ref{alg:OverView:IfCond}, the analysis checks if a given program point ({\it p\/}) is a matched program point over {\bf \ttfamily V}. If so, at line~\ref{alg:OverView:Vars}, we take \textit{p}, \textit{V} and \textit{mvicfg}, and identify all the variables alive at {\it p\/} (we also can compute invariants only for the variables users specify as interest). At line~\ref{alg:OverView:RaiseQ}, {\it RaiseQuery\/} uses the variables and the templates of interest, and instantiates an invariant candidate to form a query. A query inquires whether a particular invariant candidate holds for all the paths reachable from {\it p}. In addition, the query records the propagation in-progress information such as where the query is raised, which versions it is propagated to, the source lines that contributed to its resolution, and the node in {\it mvicfg\/} at which the query is currently located. The generated query is added to {\bf \ttfamily WorkList} at line~\ref{alg:OverView:AddWorkList}.

Once all the queries are raised, the loop at lines~\ref{alg:OverView:WhileLoop} to~\ref{alg:OverView:PropQ} propagates the query on a demand driven fashion until no queries remain in {\bf \ttfamily WorkList}. {\it PropagateQuery}, at line~\ref{alg:OverView:PropQ}, advances the queries backwards along paths in {\it mvicfg\/} until they are resolved by {\it ResolveQuery}. Whenever we have multiple paths from a node, we copy the query and propagate each instance to a path. At each edge, we perform an intersection between the versions tracked in the query and the versions marked on the edge to ensure we advance the query along a valid and interested path. For example, a query traversing versions $\left\{1,2,3\right\}$ can be propagated through an edge annotated with version $\left\{2\right\}$, and we then update the versions in the query to be $\left\{2\right\}$.

\textit{ResolveQuery} at lines~\ref{alg:ResQ} to~\ref{alg:ResolveQuery:RetFalse} takes the query, \textit{q}, the list of invariants-in-progress computed so far, {\bf\ttfamily Inv}, and {\bf \ttfamily WorkList} as input to determine whether the invariant conditions hold for the paths the query has traversed. If no decisions can be made, the query continues propagating at line~\ref{alg:OverView:PropQ}. At line~\ref{alg:ResolveQuery:IfResolved}, the {\it Resolved\/} function updates the symbolic value of the query with respect to the current node and adds the line number of the node to the query if a symbolic update has taken place. The query is resolved if (1) the constraint solver cannot find an example to show the invariant candidate fails to hold, or (2) the invariant candidates are reduced to an expression of the inputs. In the next step, \textit{CheckConflict} at line~\ref{alg:ResolveQuery:IfConflicting} compares the invariants-in-progress generated from the resolved query with {\bf\ttfamily Inv}. If {\bf\ttfamily Inv} is an empty set, we add the query to {\bf\ttfamily Inv}. When the query reports the same invariants-in-progress as any members of {\bf\ttfamily Inv}, we update {\bf\ttfamily Inv} to include the current query at line~\ref{alg:ResolveQuery:AddInv}. If any conflicts are detected, at line~\ref{alg:ResolveQuery:UpdInv}, we remove the invariants-in-progress for the conflicting versions or remove the invariants-in-progress entirely if it no longer holds for any version; we also terminate the query of the relevant versions from {\bf \ttfamily WorkList} at line~\ref{alg:ResolveQuery:UpdWL}.

At line~\ref{alg:ResolveQuery:VisitedNode}, when a query cannot be resolved, {\it VisitedNodeBefore\/} checks if any instances of the same raised query has visited the current node. If so, we check at line~\ref{alg:ResolveQuery:MergeT} whether we can merge the query. If the query has the same symbolic states as the instance cached at the node, and the versions the query propagated is the subset of the instance, we can merge the query by integrating the source lines recorded in the query to the instance. When the query has a conflicting symbolic state with the cached instance, instead of immediately terminating both the queries, the analysis compares the versions of the query with the versions stored in the conflict instance, and only terminates the versions shared by both queries. For example, when we detect a conflict between the query for program versions $\left\{2,4\right\}$ and the query for versions $\left\{2, 3\right\}$, we would no longer propagate either query over version 2 anymore. {\it MergeOrTerminateQ\/} at line~21 updates the queries in {\bf \ttfamily WorkList} based on the above merge scenarios.

When all the queries are resolved, at lines 13 to~\ref{alg:Overview:CompIC}, we process the invariants recorded in {\bf\ttfamily Inv}. The function {\it ComputingVI}, at line~\ref{alg:Overview:CompVI}, takes {\bf\ttfamily P} and {\bf\ttfamily V} to generate the versions invariants over the queried versions. To generate invariant churns, in the \textit{ComputingIC} function, we compare the invariants between consecutive program versions, generated from the same invariant candidate constructed at the same program point. We report any difference in the invariants into the invariant churns, and also report the source lines that contribute to the computations of these invariants. If needed, we can also compute invariant churns between non-consecutive versions without constructing additional queries.

\subsection{The Correctness of the Algorithm}
We believe that Algorithm~\ref{alg:OverView:GenMVICFG} computes the version invariants and invariant churns according to the definitions given in Section~\ref{sec:Overview}. First, we use MVICFGs to ensure the query is raised at the matched program points. Second, we ensure that the invariant is reported only when the condition holds for all the paths at the program point. To do so, we formulate the query using the invariant candidate, and the query resolution indicates whether the invariant candidate holds at a program point along the paths traversed by the query. When different instances of a query for the same version report a conflict, we terminate the query propagation and remove the invariant for the version. Third, version invariants are generated when for all the versions of interest (the versions don't have to be consecutive), along all the paths, the same invariant conditions hold at the program point where the query was raised. Fourth, the invariant churns are generated via differing invariants of any two versions computed from the same invariant candidates at the same program point; and thus, different invariants reported in the churns are correlated. Finally, using the MVICFG, we compare the versions marked on the edges with the versions tracked in the queries to ensure the query propagation is valid. The MVICFG makes it easy to share the query computation across different versions and also compare the invariants across versions.

\section{Evaluation}\label{sec:Evaluation}
The goals of our experiments are to demonstrate that 1) our algorithms can compute version invariants and invariant churns correctly and efficiently, and 2) version invariants and invariant churns are useful for generating assertions and locating faults  when debugging evolving programs.

\subsection{Implementation and Experimental Setup}\label{subsec:Implementation}
We implemented our algorithms in a tool called $H_2$. It uses Clang~\footnote{\url{https://clang.llvm.org/}} to compile source code and generate LLVM IR, and thus we can analyze both C and C++ code. It takes as inputs the LLVM IRs for a set of program versions as well as the source code diffs between each consecutive versions of a programs generated by the Unix diff tool to construct the MVICFG as described in~\cite{2014:ICSE:Le}. Our analyses traversed MVICFG and used Z3 constraint solver~\cite{2008:TACAS:Moura} to generate version invariants and invariant churns for a given matched program point on MVICFG\@.

We collected a total of 104 program versions from the benchmark used in~\cite{2017:FSE:Yi}. These program versions implemented 10 tasks, and each task is implemented by several different students. From each student we have an initial buggy version and then the corresponding fixed version. Such a setup can help us determine if our techniques are useful for important real-world scenarios such as patching. Meanwhile, it also helps us investigate the invariant churns and version invariants for similar code that implements the same tasks. All of the 104 program versions implemented the solutions for a simple math problem such as checking if all points are on a straight line or computing the areas of some geometric shapes. The programs are single procedures, consisting of mostly 20--50 lines of code (LOC). The documentation from the benchmark helped us easily confirm the correctness of the version invariants and invariant churns we generated.

To evaluate the computation of the version invariants and invariant churns, we selected the matched program point at the exit of the program, as it can capture real output differences and generate meaningful specifications. For each live variable $v$, we raised a query $v=?$ to inquiry if any constant invariant can exist among all versions, we report the number of version invariants generated from 6 versions of programs produced by 3 students as well as the time to compute them. Furthermore, we observe how the number of version invariants change when we increase the program versions.  To evaluate the invariant churns, we take a pair of buggy and fixed versions of a program and generate invariant churns. In addition to total invariant churns, we report how each source line diff is correlated with the changes in the program invariants. All our experimental data were collected on a Quad Intel~\textregistered\ Core~\texttrademark\ i7 CPU processor clocked
at 2.60 GHz with 16 GB of RAM running a Linux distribution.

\subsection{Results for Computing Version Invariants}\label{subsec:ExpResult}
\begin{table}[h!tb]
  \centering
  \begin{tabular}{@{\extracolsep{\fill}}ccccccc@{\extracolsep{\fill}}} \toprule
    \multirowcell{2}{\\Task ID}
                                                                                  & 2                                                      & $3$     & 4         & $5$     & 6             \\ \cmidrule{2-6}
                                                                                  & VI/T (ms)                                              & VI/T (ms) & VI/T (ms) & VI/T (ms) & VI/T (ms)     \\ \midrule
    2810                                                                          & 3/61                                                   & 3/114     & 3/161     & 3/210     & 3/240         \\ \midrule
    2811                                                                          & 6/152                                                  & 5/246     & 4/328     & 4/479     & 4/534         \\ \midrule
    2812                                                                          & 1/54                                                   & 1/66      & 1/70      & 1/110     & 1/128         \\ \midrule
    2813                                                                          & 6/192                                                  & 5/309     & 5/446     & 4/473     & 4/560         \\ \midrule
    2824                                                                          & 1/25                                                   & 1/45      & 1/58      & 1/77      & 1/84          \\ \midrule
    2825                                                                          & 5/165                                                  & 5/232     & 5/320     & 5/466     & 5/477         \\ \midrule
    2827                                                                          & 9/327                                                  & 9/451     & 9/539     & 6/715     & 6/990         \\ \midrule
    2828                                                                          & 1/91                                                   & 1/161     & 1/193     & 1/271     & 1/291         \\ \midrule
    2830                                                                          & 4/9001                                                 & 4/9056    & 4/9132    & ---/---   & ---/---       \\ \midrule
    2832                                                                          & 3/445                                                  & 3/1376    & 3/1417    & 3/1535    & 3/1574        \\ \midrule
  \end{tabular}
  \caption{Version Invariants}\label{tab:EvalVI}
\end{table}

Table~\ref{tab:EvalVI} presents the data collected for evaluating version invariants. Under \textit{Task ID}, we give the number used in the original benchmark to represent a programming task. We report the total number of version invariants (\texttt{VI}) and the time (\texttt{T}) in milliseconds required to compute them under Column \texttt{VI/T}. We collect this data from 6 program versions for each task. Among the 6 versions, versions 1 and 2 are implemented by the first student, versions 3 and 4 by the second student, and versions 5 and 6 by the third students. Versions 1, 3 and 5 are buggy versions, and versions 2, 4 and 6 are their corresponding fixed versions. Therefore, even though the task being implemented in the 6 versions is the same, the implementation differ greatly between versions 1, 3 and 5, while the changes between the buggy and fixed versions are smaller. To investigate if any meaningful version invariants potentially existent across the implementations done by different persons, we selected the versions that used the same variable names to represent the inputs and outputs. For task 2830, we only found 4 versions that satisfied this requirement.

Our results show that for each task, there exist some version invariants across all the versions analyzed. The largest number of version invariants is 6, reported for task 2827 and the minimum is 1 for 2812, 2824 and 2828.  We inspected all the reported invariants and confirmed they are indeed version invariants. This result indicates that although different persons implement a solution for the same task, the programs have ``something in common''. For example, for task 2810,  we report the version invariant for all the 6 versions, \textit{area = ($input_1$*$input_2$)/2}, where $input_1$ and $input_2$ are the two input variables. This invariant is aligned with the task description that the programs compute an area of a triangle.

We also observed that for 3 tasks, the number of version invariants decreases as we include more versions of code. The results suggest that detecting version invariants can help remove implementation specific invariants. As an example, in task \textit{2813}, we aim to find the point of intersection of two given lines. The 6 versions of the program report a total of 36 invariants. Inspecting such invariants to determine which ones are important can be time consuming~\cite{1999:ICSE:Ernst}. When we compute version invariants,  the first two versions (only differ in bug fixes) reported the invariants regarding an intermediate variable; in version 3, the program has a different way to store intermediate results, and thus the invariants regarding the intermediate variable specific to versions 1 and 2 no longer hold. Similarly, the last pair of versions had their own invariants regarding their intermediate variable, which are not common in other versions. Compared to invariants computed for a single version, version invariants are stronger and require all the versions of the same task to have such invariants. Therefore, even when a tool is unsound, detected version invariants are likely to be correct and represent important properties/specifications of a task. They are good candidates for assertions for testing or for preventing the bugs in future software evolution.

Furthermore, the change of version invariants may indicate important software property changes. For example, fixing a bug that lived in many versions of the software, or introducing a bug to a program that is well tested and always correct in the past. Sometimes, the change of version invariants inform an algorithm change. In our experiment, task \textit{2827} aims to determine if all three given points fall in a straight line. The first four versions used an algorithm  that calculated the distance to determine the outcome; however, the last two versions used the slope to calculate the outcome. Thus, the number of version invariants is reduced when including version 5.

We have collected the time used to compute the version invariants. The analysis can finish in seconds for all the cases. We observe that the time required to compute the version invariants increased only marginally between a buggy and its corresponding fixed version with an average increase of 53.8 ms. This demonstrate the advantage of using MVICFG to compute version invariants over traditional method of computing them individually and then comparing them. When versions with different implementations were added, the time required to compute the version invariants increased by an average of 1.48 s as the versions do not share as much source code. When computing version invariants for an additional version, such performance difference provides hints on whether we introduced a similar version or a quite different version.

\subsection{Results for Computing Invariant Churns}\label{subsubsec:EvalIC}
\begin{table}[h!tb]
  \centering
  \resizebox{\columnwidth}{!}{%
    \begin{tabular}{@{\extracolsep{\fill}}cccccccc@{\extracolsep{\fill}}} \toprule
      Pair ID & CChurn & LOC & IChurn & Inv & Fix & IChurn/line & T (ms)     \\ \midrule
      270025  & 1      & 2   & 1      & 10  & Y   & 1           & 58        \\ \midrule
      270048  & 1      & 2   & 1      & 6   & Y   & 1           & 31        \\ \midrule
      270053  & 1      & 4   & 1      & 5   & Y   & 0.25        & 49        \\ \midrule
      270057  & 1      & 2   & 1      & 5   & Y   & 0           & 48        \\ \midrule
      270142  & 1      & 2   & 6      & 6   & Y   & 3           & 184       \\ \midrule
      270150  & 2      & 4   & 2      & 12  & Y   & 1.5         & 96        \\ \midrule
      270194  & 2      & 7   & 2      & 10  & Y   & 0.14        & 934       \\ \midrule
      270201  & 1      & 2   & 3      & 14  & Y   & 3           & 90        \\ \midrule
      270220  & 1      & 6   & 5      & 18  & Y   & 4           & 115       \\ \midrule
      270227  & 3      & 12  & 4      & 4   & Y   & 0.33        & 114       \\ \midrule
      270330  & 1      & 10  & 0      & 2   & N   & 0           & 15        \\ \midrule
      270357  & 3      & 10  & 1      & 1   & Y   & 0.09        & 46        \\ \midrule
      270391  & 1      & 2   & 0      & 12  & N   & 0           & 75        \\ \midrule
      270406  & 1      & 2   & 6      & 6   & Y   & 3           & 58        \\ \midrule
      270466  & 1      & 4   & 2      & 16  & Y   & 1.5         & 134       \\ \midrule
      271056  & 1      & 1   & 4      & 4   & Y   & 4           & 24        \\ \midrule
      271154  & 1      & 2   & 1      & 12  & Y   & 0.5         & 72        \\ \midrule
      271206  & 1      & 5   & 6      & 6   & Y   & 1.4         & 194       \\ \midrule
      271340  & 1      & 2   & 10     & 10  & Y   & 5           & 139       \\ \midrule
      271410  & 1      & 2   & 1      & 2   & Y   & 0.5         & 12        \\ \midrule
      271464  & 1      & 2   & 1      & 2   & Y   & 0.5         & 15        \\ \midrule
      271471  & 1      & 2   & 1      & 1   & Y   & 0.5         & 8         \\ \midrule
      271876  & 2      & 2   & 1      & 7   & Y   & 0.5         & 47        \\ \midrule
    \end{tabular}
  }
  \caption{Invariant Churns}\label{tab:EvalIC}
\end{table}

We took all the buggy/fixed versions from the programs used for calculating version invariants and removed the pairs where the changes only effect the output strings, e.g., ``adding a period at the end of the string''. As a result, we collected 23 pairs of programs for this experiment. We show our results in Table~\ref{tab:EvalIC}. Under {\it Pair ID}, we give the number used in the original benchmark to represent these pairs.  The column \textit{CChurn} reports the total number of source code churns between the two versions, and \textit{LOC} shows the total size of the code churns in terms of LOC\@. Under {\it IChurn}, we show the number of invariant churns. As a comparison, we detect program invariants for each individual version, and provide the total number in Column {\it Inv}.  Under {\it Fix}, we list whether the invariant churn reflects the effectiveness of a patch. Since these programs all performed mathematical computations, and the errors and patches are all related to math formulas, we mark ``Y'' where the churn demonstrates the removal of the buggy formula and the addition of the correct formula. Under \texttt{IChurn/line} we report the average number of invariant churns per source code churn line. That is, if a developer provides a source line involved in the diff, and she/he wants to know how it impacts the semantics of the program, this is the number of invariant churns she/he will inspect. Finally, under \texttt{T (ms)}, we present the total time in ms taken for computing the invariant churns.

The columns {\it CC\/} and {\it LOC\/} indicate that the patches include 1--3 churns with the size from 1--13 LOC\@. The results also show that the number of invariant churn reported varied between 1 and 10 with an average of 2.34 invariant churn per program. The majority of the pairs report 0--2 invariant churns, compared to Column {\it Inv}, where close to half of the program pairs report 10 or more invariants. That means that with our tool, the developers do not need to manually align these invariants to determine which pairs are related to the changes. There are also pairs where the total number of invariants and the number of invariant churns are the same. In the buggy versions of these programs, none of the variables were initialized properly, resulting in the change of all the invariants for the fixed version. This indicated a situation in which all the variables present in a program, changed between versions. If we omit these cases, there are on average 1.64 invariant churn reported per program versions versus 7.43 total invariants per program versions. We manually inspected all the invariant churns and found our tool correctly computes all the invariant churns.

Our inspection also determines that there are 21 out of 23 pairs where the invariant churns report the change from a buggy condition to a correct condition, among which 10 of them are the only churns reported, i.e., no false positives. As an example, in \textit{270025}, we calculate the area of the triangle formed by the three points (a,b), (a,0) and (0,b). Buggy version reported the area as a negative number in some cases. The fixed version corrected this problem by calling the \texttt{abs} function when calculating the area. The only invariant churn for \textit{270025} $\{-area = (0.5 * input_1 * input_2), +area = (0.5 * (ite ((input_1 * input_2 ) >= 0) (input_1 * input_2) (-(input_1 * input_2))))\}$ where $input_1$ and $input_2$ are the unique input points (here the invariants are reported in the Z3 constraint format). It indicates that in the patched version, the area has to be always positive while in the buggy version, it can be negative. The two cases in which the invariant churns were not able to capture such important change semantics was because the fix was in the control flow related to how the result is returned but not in the computed results themselves. In these cases, the invariants associated with the output variables didn't change, but the condition leading to their calculation changed. Such conditions are not recorded in a variable, and thus not included in the invariant candidate. We observed that in all the 21 cases, the second invariants from bug fixes can be used to generate assertions.

We also found that invariant churns can be used in debugging to quickly locate the bugs since they record the source code lines contributing to the change of invariants as well. Consider a scenario in which a source code change broke a functionality in a program. The invariant churns will highlight only the invariants that changed instead of having to compare all the invariants across both the versions. If we are able to find the invariant churn causing the regression, it will contain the source code lines contributing to the change which can narrow down the faulty location. For example in \textit{270150}, where the task was to determine the point of intersection of two lines, the difference between the fixed and buggy code is a change in the arithmetic formula. There are 4 source lines of code changed between the buggy and fixed version and 12 invariants to compare between the versions but only 2 invariant churns. One of the invariant churn $\{-X = ((input_2 - input_4)/((input_3 * input_2) - (input_1 * input_4))), +X = ((input_4 - input_2)/((input_1 * input_4) - (input_3 * input_2)))\}, \{\{\pm7,\pm6\}, \emptyset\}  $, where $input_1$ to $input_4$ are unique input points, capture the change in the formula as well as the lines contributing to the change. From this invariant churn we can see how the invariants changed when we corrected the formula used for computation of variable $X$ as well as the fact that the change in source code lines 6 and 7, contributed to it.

On the other hand, if we only knew the source lines that changed, but not the problematic invariant changes, we can use the lines associated with the invariant churn to narrow down the possible culprits. For example, in \textit{270201}, which was implementing the same task as \textit{270150}, we only changed one source line, but the variable we modified was used by other variables resulting in changes else where in the program with 3 invariant churns being reported. If we filter the invariant churns based on the line we modified, we can easily see that the line contributed to all 3 of the invariant churns.  Under {\it IChurn/line}, we report the number of invariant churns per source line. The average number of invariant churns per line is 1.37 with a range between 0 and 4. It indicates that for the majority of the cases, given a changed line of interest, we only need to inspect less than 2 invariant churn to determine the impact of the lines. Since this number is small and the invariant churns can capture the effect of the source code change, it may be helpful to report invariant churns as a way of describing bug fixes or the impact of code change during a code review session.

Under {\it T (ms)}, we show that the time required to compute the invariant churns is less than a second with an average of 111.21 ms per program.

In summary, this experiments shown that we can compute invariant churns correctly in reasonable amount of time and that it is able to capture the semantics of the bug fix concisely as opposed to computing every invariant and checking for changes between them. Our results also suggest that invariant churn might be beneficial auxiliary information during a code review to help understand and verify changes.

\subsection{Limitations}
$H_2$ is a prototype to evaluate the correctness of the Algorithm 1 presented in Section 3. Our current implementation performs an intra-procedural analysis and handles a loop by unrolling it once. We only applied pointer analysis provided by LLVM\@. The tool supported a set of instructions related to mathematical computations. Due to the above reasons, we may compute incorrect invariant diffs. The benchmark we used in the experiments solve mathematical problems, and thus we have not encountered incorrect invariant diffs in our results. In our evaluation, we used the equality invariant templates. We may generate more invariant diffs and incur more time if more templates are applied. Thus the numbers reported in the tables are tied to the configurations we used to run the tool. Using the benchmark programs, we have demonstrated the usefulness of the version invariants and invariant churns. We believe that these use scenarios are applicable to real-world programs, but we may discover additional application scenarios if we further study the real-world software repositories.

\section{Related Works}\label{sec:Related}
The idea of representing program properties using invariants have been explored extensively in a variety of fields including specification mining~\cite{2002:POPL:Ammons,2007:ICSE:Ramanathan,2007:PLDI:Ramanathan}, program verification~\cite{2012:ICSE:Ghardallou,2008:HASE:Jicheng,2004:ICTAC:Rodriguez-Carbonell,2005:SYNASC:Kovacs}, software testing~\cite{2015:ICSE:Kim,2013:ASE:Sagdeo}, and program refactoring~\cite{2001:ICSM:Kataoka,2007:EASST:Massoni}. There also have been works on representing semantic changes using conditions of program variables~\cite{2012:FSE:Qi,2015:ACM:Yi,2013:ISSTA:Yi,2014:ICSME:Le,2014:OOPSLA:Partush,2013:FSE:Lahiri,2012:CAV:Lahiri,2008:SIGSOFT:Person,2017:ISSTA:Gyori}. The works such as change contracts and differential assertions represent the states of a new program using the states of an old program. Whereas, invariant diffs report the commonalities and changes using the program variables from their own versions, and the connections of the program versions are established through MVICFG using the matched program points, shared nodes and shared paths.

Since Notkin~\cite{2002:LPA:Notkin} proposed the idea of analyzing evolving programs longitudinally to reduce the cost of program analysis, most of the work have focused on using semantic equivalence~\cite{2008:SIGSOFT:Person,2012:CAV:Lahiri} to carry the intermediate analysis results from one version to another. While it is able to reuse some computation across versions, they still work with each version separately. Another approach used by Partush et al~\cite{2014:OOPSLA:Partush} involves creating an unified version of the program by merging two programs using an abstract domain. As the abstraction loses information, their approaches can generate false positives. Our approach analyzed the shared nodes for multiple versions only once via MVICFG\@. The sharing does not sacrifice the precision. We can compare the intermediate results computed for different versions, and terminate the version invariant detection early when conflicted invariant conditions  from different versions and paths are detected.

Similar to other demand-driven approaches~\cite{2008:FSE:Le,2005:OOPSLA:Sridharan,2001:PLDI:Heintze,2011:ISSTA:Yan,2008:POPL:Zheng,2007:PASTE:Le,2014:ICSE:Le}, we formulate our problem, of finding invariant diffs, as computing query resolutions. A small difference is that since we are seeking for invariants which have to hold along all the paths, our demand driven analysis returns the invariants when all the queries are resolved.

Using invariants to generate assertions and help localize the faults is a well explored area with works like~\cite{2010:DATE:Vasudevan,2013:GLSVLSI:Lin,2002:SIGSOFT:Nimmer,2001:TCS:Nimmer,2009:ASE:Daniel}. Since we can compute invariants that have been present in multiple versions of a program, we may provide better guarantee of the assertion generated being true. There are also works~\cite{2013:ASE:Sagdeo,2013:SIGPLAN:Sahoo} that computes likely invariant to localize the faults for a program. The goal of these works is to narrow down the set of suspicious statements with the help of invariants. Our work fits the scenarios of diagnosing faults introduced in a change. We statically compute a set of invariant changes to help understand the effect of changes.

\section{Conclusions}\label{sec:Conclusion}
This paper presents invariant diffs as a way to specify changes between program versions for continuously evolving programs. We designed a static demand-driven, path-sensitive analysis on the MVICFG to compute such invariant diffs. The key impact of invariant diffs is twofold. First, version invariants can filter some implementation specific conditions, and the common properties of a set of versions may more likely represent meaningful specifications. We compute version invariants by analyzing multiple versions of programs simultaneously using MVICFG, avoiding redundant computation. Second, the invariant churns captures the effect of the code change and correlates it to the code churn. It enables us to pinpoint the changed program properties over two versions given a source code churn or even a line in the code churn. Similarly, we are also able to identify the part of the source code churn that is responsible for the change of program properties. We demonstrate that our algorithm is able to correctly compute the version invariants and invariant churns within reasonable amount of time. We show that the version invariants can represent the specifications of a common task implemented by different persons, and the invariant churns can represent the semantics of the bug fix. We also found that the number of invariant churns is more manageable compared to the number of program invariants, and thus there is a great potential to use invariant churns to assist code review for changes. In the future, we will further explore the scalability and usability of invariant diffs for real-world software repositories.

\section*{Acknowledgement}
This research is supported in part by the US National Science Foundation (NSF) under Award 1542117.

\bibliographystyle{IEEEtran}
\bibliography{ms}
\end{document}